\documentclass[prb,showpacs,preprintnumbers,twocolumn,amsmath,amssymb]{revtex4}
\usepackage{graphicx}
\usepackage{dcolumn}
\usepackage{bm}

\newcommand{\PrOsSb}{PrOs$_4$Sb$_{12}$}
\newcommand{\PrRuSb}{PrRu$_4$Sb$_{12}$}
\newcommand{\PrOsRuSb}{Pr(Os$_{1-x}$Ru$_x$)$_4$Sb$_{12}$}
\newcommand{\UPt}{UPt$_{3}$}
\newcommand{\UPtPd}{U(Pt$_{1-x}$Pd$_x$)$_{3}$}
\newcommand{\UThBe}{U$_{1-x}$Th$_{x}$Be$_{13}$}

\begin{document}

\title{Double superconducting transition and low-temperature specific
heat study of \PrOsRuSb}

\author{N. A. Frederick}

\author{T. A. Sayles}

\author{S. K. Kim}

\author{M. B. Maple}

\affiliation{%
Department of Physics and Institute for Pure and Applied Physical
Sciences, University of California at San Diego, La Jolla, CA
92093
}%

\date{\today}

\begin{abstract}

The double superconducting transition in \PrOsSb, first observed
in specific heat $C(T)$ measurements of single crystal samples, is
the subject of an in-depth study. The double superconducting
transitions of a batch of single crystals were measured before and
after they were annealed for $5$ days at $500^{\circ}$C, with no
observed change. $C(T)$ measurements near $T_c$ for \PrOsSb{} in
several magnetic fields are also presented, detailing the
evolution of the double transition up to $1$ T. Samples of
\PrOsRuSb{} with $0.01 \leq x \leq 0.04$ also appear to display a
double superconducting transition in specific heat. In addition,
samples with the smallest Ru concentration measured ($x=0.01$) may
even display a different type of superconductivity than pure
\PrOsSb{} ($x=0)$.

\end{abstract}

\pacs{65.40.Ba, 71.27.+a, 74.25.Bt, 74.62.Dh}

\keywords{}

\maketitle

\section{Introduction}

The compound \PrOsSb{} has attracted a great deal of attention in
recent years as the first Pr-based heavy fermion superconductor,
with a superconducting transition temperature $T_c = 1.85$ K and
an electron effective mass $m^* \approx 50
m_e$.\cite{Bauer02a,Maple02a} Detailed measurements show the
strongest support for a nonmagnetic $\Gamma_1$ singlet ground
state in a tetrahedral crystalline electric field (CEF), separated
from a $\Gamma_5$ (in the cubic notation) triplet first excited
state by approximately $10$
K.\cite{Kohgi03,Rotundu04a,Goremychkin04} Quadrupolar ordering
arises below $1.5$ K and above $4.5$ T, related to the crossing of
the Zeeman-split CEF energy levels.\cite{Maple03a,Ho03,Kohgi03}
The superconducting state appears to be highly unconventional,
exhibiting time-reversal symmetry breaking,\cite{Aoki03} multiple
superconducting phases,\cite{Izawa03,Cichorek05} and double
superconducting
transitions.\cite{Maple02a,Oeschler03,Vollmer03,Tayama03,Chia03}

Upon substitution of Ru to form \PrOsRuSb, the transition
temperature decreases to a minimum of $T_c \approx 0.75$ K at
$x=0.6$, whereupon it increases to $\sim 1.0$ K for \PrRuSb.  This
behavior was observed in measurements of electrical resistivity,
magnetic susceptibility, and specific
heat.\cite{Frederick04,Frederick05}  All three measurement
techniques indicated that the CEF energy level splitting between
the ground state and first excited state increased monotonically
from $\sim 10$ K for \PrOsSb{} to $\sim 70$ K for \PrRuSb.  The
$C(T)$ experiments also revealed that the electronic specific heat
coefficient $\gamma$ decreased from $\sim 500$ mJ/mol K$^2$ for
$x=0$ to a plateau of $\gamma \sim 100$ mJ/mol K$^2$ for $x \geq
0.6$.  Upon subtracting the normal state $\gamma$ and Schottky
anomaly due to CEF effects, we were able to investigate the
structure of $C(T)$ below $T_c$.  We performed phenomenological
fits which suggested that the \PrOsSb{} data below $T_c$ were more
in agreement with power law behavior, indicating nodes in the
energy gap, while the samples with $0.05 \leq x \leq 0.2$
suggested exponential behavior, consistent with an isotropic
energy gap.  In addition, no obvious double superconducting
transition was observed for $x=0.05$, the lowest substituted
concentration measured in the previous study.

In this paper, we present the results of further probes into the
nature of the double superconducting transition in \PrOsSb.  We
observe the effect of magnetic fields, annealing, and Ru
substitution on the double transition through measurements of
$C(T)$.  These results are compared to the double superconducting
transitions observed in the heavy fermion systems \UPtPd{} and
\UThBe.  The low-temperature specific heat of \PrOsRuSb{} is
investigated in order to determine a possible concentration at
which the superconducting state changes from one possessing nodes
in the energy gap to one that is more isotropic.

\section{Experimental Details}

The single crystals of \PrOsRuSb{} studied in this work were grown
using an Sb flux method, and the crystal structure was confirmed
via x-ray diffraction measurements, as described previously.
\cite{Frederick04} The specific heat $C(T)$ measurements were
performed in the same cryostat as for previous experiments,
\cite{Frederick05} using a semi-adiabatic method at temperatures
between $0.6$ K and $20$ K. Magnetic fields were applied using a
$5$ T superconducting magnet attached to the outside of the vacuum
can. The experiments in a magnetic field were performed on the
same individual single crystal measured by Cichorek {\it et
al.},\cite{Cichorek05} with a mass of $8.17$ mg.  The field
direction was along a principal crystal axis.  The other
measurements were made on collections of single crystals, with
total masses between $20$ and $30$ mg.

The annealing was carried out by wrapping the crystals in tantalum
foil, which was sealed inside a quartz tube with a piece of
zirconium foil, under $150$ Torr UHP Ar.  The quartz tube was
placed in a $500~^{\circ}$C furnace for $5$ days, and then
quenched to room temperature.  The mass of the sample was the same
after annealing within the systematic error of the weighing
scales.

\section{Results and Discussion}

\begin{figure}[tbp]
\begin{center}
\includegraphics[angle=270,width=3.375in]{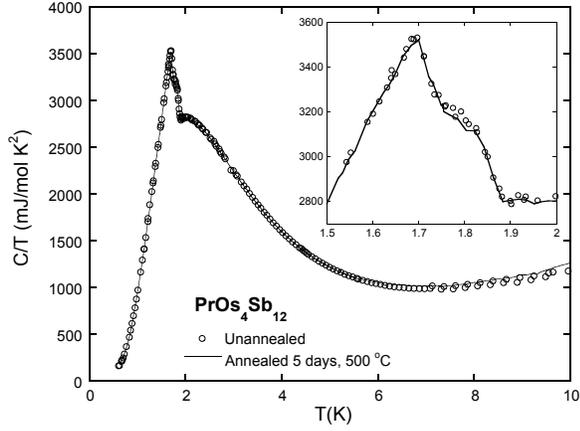}
\end{center}
\caption{Specific heat divided by temperature $C/T$ versus $T$
below $10$ K for an unannealed set of single crystals of \PrOsSb{}
(open circles) and the same sample annealed for $5$ days at
$500~^{\circ}$C (solid line). Inset: close-up of the double
superconducting transition. There is almost no difference between
the two data sets.} \label{Annealing}
\end{figure}

The only other {\it stoichiometric} heavy fermion superconductor
that displays a double superconducting transition is
\UPt.\cite{Fisher89,Joynt02} For single crystals of this compound,
annealing can be an important factor in observing the double
transition in specific heat.\cite{Kim97} The results of an
annealing study on a batch of \PrOsSb{} crystals are shown in
Fig.~\ref{Annealing}. The two data sets fall neatly on top of each
other, which argues against any kind of inhomogeneous mass
redistribution. Unlike \UPt, however, the double transition in
\PrOsSb{} shows no indication of becoming more resolved due to
annealing.  It is still important to note that the double
transition does not degrade with annealing, as might be expected
if this feature were not an intrinsic effect of \PrOsSb.

\begin{figure}[tbp]
\begin{center}
\includegraphics[angle=270,width=3.375in]{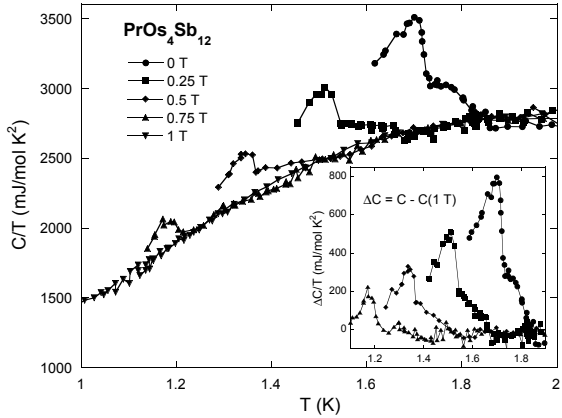}
\end{center}
\caption{Specific heat divided by temperature $C/T$ versus $T$
between $1$ and $2$ K in magnetic fields up to $1$ T for an
individual single crystal sample of \PrOsSb.  The data for fields
below $1$ T have been truncated below the superconducting
transitions for clarity. Inset: $\Delta{}C/T$ versus $T$, where
$\Delta{}C = C(H,T) - C(H = 1 {\rm T}, T)$.} \label{Fields}
\end{figure}

Figure~\ref{Fields} displays the evolution of the double
superconducting transition with magnetic fields for \PrOsSb.  The
double transition is clearly visible in the zero field data, while
by $1$ T the transition has been suppressed below $1$ K.  As the
magnetic field is increased, the first transition appears to lose
its sharpness, until by $0.75$ T it is either nonexistent, or
indistinguishable within the strong downturn of the Schottky
anomaly.  The inset to Fig.~\ref{Fields} shows the data after
subtracting the $1$ T curve, which further emphasizes the initial
jumps.  This evolution of the transition with magnetic field has
been reported by several other groups, with mostly different
results.  Vollmer {\it et al.}~found that the double transition
disappears by $0.4$ T,\cite{Vollmer03} while the double transition
observed by Grube {\it et al.}~persists up to $0.6$
T,\cite{Grube05} similar to this work. The data reported by
Measson {\it et al.}~are noticeably different, as two clear
transitions were visible up to $\sim 2$ T.\cite{Measson04}  All
results, except those of Vollmer {\it et al.}, were from
experiments performed on one single crystal of \PrOsSb; the
experiments of Measson {\it et al.}~relied on an ac calorimetry
method instead of the semi-adiabatic method utilized by Vollmer
{\it et al.}, Grube {\it et al.}, and this work. In \UPt, applying
a magnetic field and lowering the temperature causes the two
transitions to merge near $\sim 0.5$ T and $\sim 0.4$ K, depending
on the field orientation, giving rise to an additional phase
transition within the superconducting
state.\cite{Hasselbach89,Lohneysen94} While several experiments
provide evidence for multiple phases in the superconducting state
of \PrOsSb, none of these phases can as yet be associated with the
double superconducting transition.\cite{Izawa03,Cichorek05}

\begin{figure}[tbp]
\begin{center}
\includegraphics[angle=270,width=3.375in]{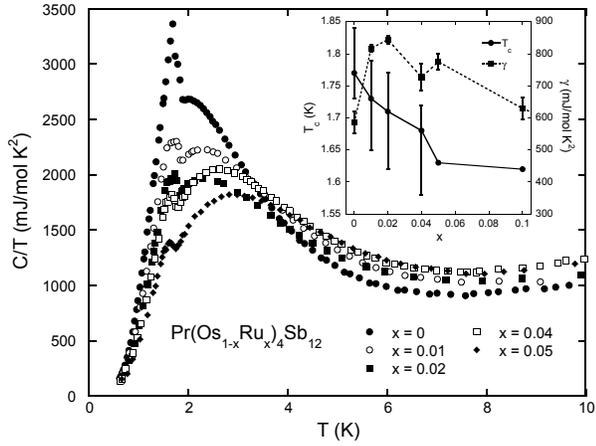}
\end{center}
\caption{Specific heat divided by temperature $C/T$ versus $T$
below $10$ K for \PrOsRuSb{} with $0 \leq x \leq 0.05$.  Inset:
superconducting transition temperature $T_c$ (filled circles, left
axis) and the electronic specific heat coefficient $\gamma$
(filled squares, right axis) as a function of Ru concentration
$x$.  The vertical lines for $T_c$ delineate the upper and lower
transitions for those concentrations which display double
transitions.} \label{lowxC}
\end{figure}

Specific heat divided by temperature $C(T)/T$ data for \PrOsRuSb{}
with $0 \leq x \leq 0.05$ are displayed in Fig.~\ref{lowxC}.  The
normal state data below $10$ K were fitted with an equation that
includes electronic, lattice, and CEF contributions, with the
results listed in Table \ref{Table1}.  The quantity $r$ is a
scaling factor for the Schottky anomaly, as discussed
previously.\cite{Frederick05}  This suppression of the Schottky
anomaly could result from an energy dispersion due to Pr-Pr
interactions or hybridization between the Pr f-electrons and
ligand states, which results in entropy being transferred to the
conduction electrons, thereby increasing the electronic specific
heat coefficient $\gamma$.  The errors for the parameters were
determined by varying the Debye temperature $\Theta_{\rm D}$ by
$\pm 10$ K from its best fit value, as described
previously.\cite{Frederick05} Only results from a $\Gamma_1$
ground state fit are presented here; we were able to accurately
use the simpler cubic CEF equations since these measurements were
only performed in zero field.  As can be seen from Table
\ref{Table1}, the splitting between the ground state and first
excited state, $\delta$, increases monotonically with increasing
$x$, as expected.  The electronic specific heat coefficient
$\gamma$, on the other hand, displays a maximum at $x=0.02$. The
Debye temperature $\Theta_{\rm D}$ is also enhanced for $x=0.01$
and $x=0.02$ compared to the neighboring concentrations.

\begin{figure}[tbp]
\begin{center}
\includegraphics[angle=270,width=3.375in]{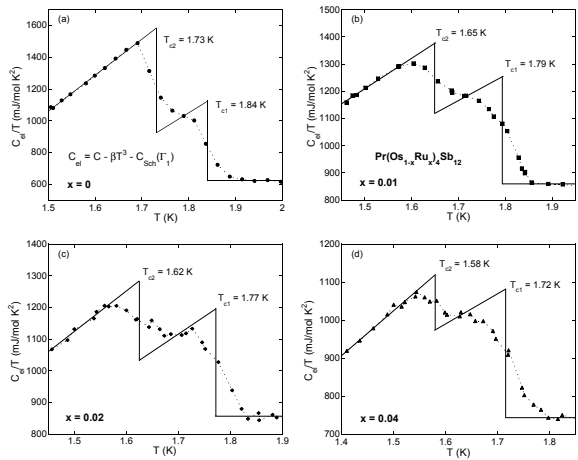}
\end{center}
\caption{Comparison of double superconducting transitions in
$C_{\rm el}/T$, after lattice and Schottky anomaly terms
corresponding to a $\Gamma_{1}$ ground state have been subtracted,
for \PrOsRuSb{} with (a) $x=0$, (b) $x=0.01$, (c) $x=0.02$, and
(d) $x=0.04$.  The transitions were approximated with an entropy
conserving construction.} \label{Transitions}
\end{figure}

\begin{figure}[tbp]
\begin{center}
\includegraphics[angle=270,width=3.375in]{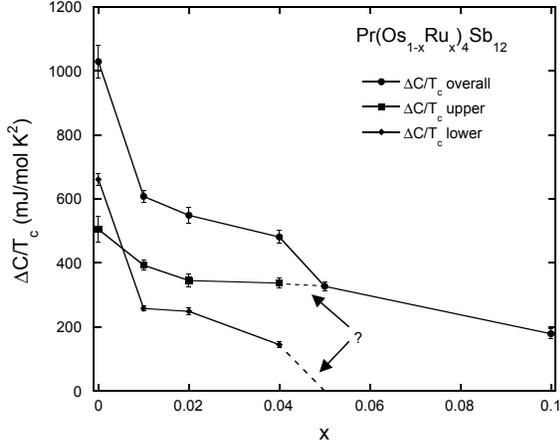}
\end{center}
\caption{Specific heat jump at $T_c$, $\Delta{}C/T_c$, for
\PrOsSb{} as a function of Ru concentration $x$.  The dashed lines
represent speculations on the continued evolution of
$\Delta{}C/T_c$.} \label{DeltaCPhase}
\end{figure}

The superconducting transitions in Fig.~\ref{Transitions} are
plotted as $C_{\rm el}(T)/T$, where $C_{\rm el}$ is the electronic
specific heat after the lattice and Schottky terms were
subtracted. Structure in the superconducting transition is clearly
apparent for $x=0.01$ and $x=0.02$, and also appears to be present
for $x=0.04$.  In light of these data, the rounded structure of
the $x=0.05$ transition published previously\cite{Frederick05}
might also hide a double transition, but it is too broad to tell
for sure.  The transitions for $0 \leq x \leq 0.04$ were analyzed
using an entropy conserving construction for both transitions. The
upper and lower transitions and their respective specific heat
jumps are presented in Table \ref{Table2}.  It is interesting to
note that the lower transition has the largest jump for \PrOsSb,
while for $0.01 \leq x \leq 0.04$, the upper transition displays a
larger jump.  This is shown graphically in Fig.~\ref{DeltaCPhase},
which allows us to speculate that the upper transition jump and
the overall transition jump may converge near $x=0.05$.  Also
listed in Table \ref{Table2} is the ratio
$\Delta{}C/\gamma{}T_{c}$ for the full transition.  The value for
$x=0$ of $1.76$ is enhanced over the BCS value of $1.43$,
suggesting strong-coupling superconductivity.  The results for the
upper and lower transitions are not tabulated, although they can
be easily calculated from the other supplied data.  The upper and
lower transitions have $\Delta{}C/\gamma{}T_{c}$ values of $0.86$
and $1.12$, respectively.  As $x$ increases,
$\Delta{}C/\gamma{}T_{c}$ drops sharply, due to the strong
increase in $\gamma$ and the smaller decrease in $T_c$, to values
well below $1.43$, even for the full transitions.

The evolution of the double transition in \PrOsRuSb{} is
qualitatively similar to that seen in \UPtPd, which displays a
double transition down to $x = 0.003$, but with no
superconductivity observed at all for $x > 0.004$ above $1$
K.\cite{Keizer99}  The lower transition is very small in \UPtPd{}
for $x=0.003$, but still may be present for higher concentrations
at temperatures lower than what was measured.  Nevertheless, the
specific heat jump at the lower transition in both series of
compounds decreases in magnitude faster than the jump for the
upper transition.  An interesting discrepancy between the two
substituted series is that in \UPtPd, the difference between the
two superconducting transitions, $\Delta{}T_c$, increases with
increasing $x$, while in \PrOsRuSb{} the separation between the
transitions remains nearly constant. Unfortunately, the smaller
amount of research on \PrOsSb{} makes it difficult to determine
which similarities and differences between \PrOsRuSb{} and
\UPtPd{} are meaningful. It should be noted that a double
superconducting transition has been reported for
Pr$_{1-x}$La$_{x}$Os$_4$Sb$_{12}$ with $x = 0.02$ and $x = 0.05$,
and the lower transition also appears to be the one suppressed
with La substitution.\cite{Rotundu04b}

\begin{figure}[tbp]
\begin{center}
\includegraphics[angle=270,width=3.375in]{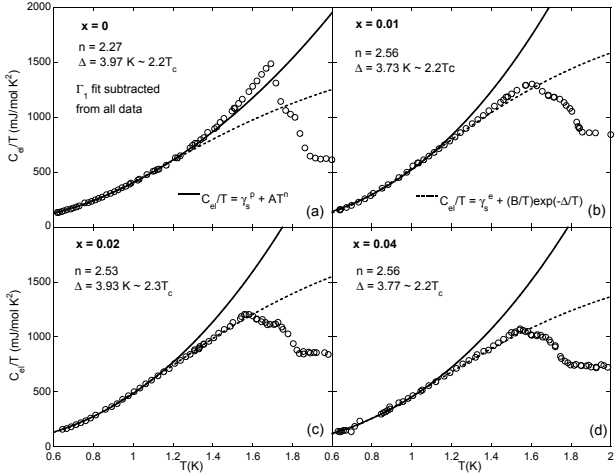}
\end{center}
\caption{Comparison of exponential (dashed line) and power law
(solid line) fits below $T_c$, after subtracting lattice and
Schottky anomaly terms for a $\Gamma_{1}$ ground state, for
\PrOsRuSb{} with (a) $x=0$, (b) $x=0.01$, (c) $x=0.02$, and (d)
$x=0.04$. The $x=0$ fits only extend up to $\sim 1.2$ K, while the
$x=0.01$, $x=0.02$. and $x=0.04$ fits only extend up to $\sim
1.15$ K, as described in the text.} \label{SCcompare}
\end{figure}

Figure~\ref{SCcompare} displays fits to the data below $T_c$ for
both power law (appropriate for energy gaps with nodes) and
exponential (appropriate for isotropic energy gaps) functions, a
continuation of the fits performed previously.\cite{Frederick05}
The new fits for $x=0.01, 0.02$, and $0.04$ were all made between
$\sim 0.6$ K (the base temperature of the measurement) and $1.15$
K. Neither fit is better than the other within the fit range, but
the phenomenological extrapolation of the fits to higher
temperatures suggests that there may be a fundamental difference
between the superconductivity in \PrOsSb{} and the Ru-doped
materials, even with only $1\%$ Ru.  A specific heat study of
\UThBe{} deep in the superconducting state revealed a similar
evolution from power-law to BCS-like behavior with increasing $x$,
separated at $x \approx 0.02$.\cite{Jin94}  The parameters from
fits to the \PrOsRuSb{} superconducting state data are listed in
Table \ref{Table2}. A rough estimate for the coupling strength can
be obtained by calculating $2\Delta_{e}/T_{c}$ (with $\Delta_{e}$
in units of K), which is approximately $4.5$ for the values of $x$
studied in this work. This enhancement over the BCS estimate of
$3.52$ is consistent with other measurements on \PrOsSb{} which
indicate a strong-coupling energy
gap.\cite{MacLaughlin02,Kotegawa03}

A double superconducting transition is one of the clearest
indications of unconventional superconductivity, and possibly a
multi-component superconducting order parameter. Studies of \UPt{}
strongly indicate that the lower transition in this compound
arises from the onset of a new order parameter (presumably
antiferromagnetic);\cite{Joynt02} the double transition in
\PrOsSb{} is not, as yet, as well studied or understood.  Time
reversal symmetry breaking (TRSB) has been observed in \PrOsSb{}
through measurements of muon spin rotation ($\mu$SR) in zero
field, which exhibit a spontaneous magnetic moment arising in the
superconducting state.\cite{Aoki03} It is difficult to tell from
the published data whether the moment arises due to the upper or
the lower superconducting transition. In \UPt, a spontaneous
moment in $\mu$SR was reported below the lower superconducting
transition,\cite{Luke93} but this result has not proven to be
reproducible via $\mu$SR
measurements.\cite{deReotier95,Higemoto00} However, ultrasound,
vortex, and flux flow measurements have revealed features below
the lower superconducting transition associated with additional
superconducting order parameters and, possibly, with
TRSB.\cite{Ellman96,Amann98,Shung98,Joynt02} A magnetic moment has
been observed to arise in $\mu$SR measurements below the lower
superconducting transition of \UThBe{} with $0.02 \leq x \leq
0.04$, also suggesting TRSB.\cite{Heffner90} Measurements of
vortex motion in \UThBe{} with $x = 0.0275$ below the lower
superconducting transition indicate a sharp drop in vortex creep
rates, very similar to that seen in \UPt.\cite{Dumont02}  It
therefore would be very useful to perform these measurements on
low $x$ samples of \PrOsRuSb{} in order to determine if the TRSB
and unconventional superconductivity truly track the double
superconducting transition, as these present measurements suggest.

\section{Summary}

In summary, we have investigated the double superconducting
transition in \PrOsSb{} through specific heat measurements in more
detail than previously reported. Annealing the crystals did not
appear to affect the features in the specific heat near $T_c$.  In
a magnetic field, the transitions rapidly broaden in temperature
and are suppressed by $1$ T. Samples of \PrOsRuSb{} with smaller
Ru concentrations than previously measured displayed double
superconducting transitions in $C(T)$ up to $x=0.04$.  Power law
and exponential fits to $C(T)$ below $T_c$ also suggest that the
superconducting state for $x=0$ is different than that for $x \geq
0.01$.  The behavior of the double superconducting transition for
\PrOsRuSb{} appears to only share qualitative characteristics with
the superconducting heavy fermion systems \UPtPd{} and \UThBe.

\section*{Acknowledgements}

We would like to thank W. M. Yuhasz for experimental assistance.
This research was supported by the U.S. Department of Energy Grant
No.~DE-FG02-04ER-46105 and the U.S. National Science Foundation
Grant No.~DMR-03-35173.

\newpage

\begin{table}[p]
\caption{Normal state physical properties of samples of \PrOsRuSb,
determined from specific heat data. The parameter $\delta$ is the
splitting between the ground state and the first excited state in
the Schottky anomaly, $r$ is the scaling factor for the Schottky
anomaly (as described in the text), $\gamma$ is the estimated
electronic specific heat coefficient, and $\Theta_{D}$ is the
estimated Debye temperature. The errors in the parameters were
determined by allowing $\Theta_{D}$ to vary by $\pm10$ K within
the fits (see text and Ref.~\cite{Frederick05} for
details).}\label{Table1}
\bigskip
\begin{tabular}{|c||cccc|}
\hline $x$ & $\delta$ & $r$ & $\gamma$ & $\Theta_{D}$ \\
 & (K) & & (mJ/mol K$^{2}$) & (K)\\
\hline
0    & 7.36$\pm$0.04 & 0.56$\pm$0.01 & 586$\pm$33 & 211 \\
0.01 & 8.28$\pm$0.01 & 0.43$\pm$0.01 & 817$\pm$10 & 278 \\
0.02 & 8.43$\pm$0.01 & 0.35$\pm$0.01 & 844$\pm$12 & 271 \\
0.04 & 9.19$\pm$0.01 & 0.45$\pm$0.02 & 727$\pm$42 & 201 \\
0.05 & 10.2$\pm$0.01 & 0.39$\pm$0.01 & 775$\pm$25 & 224 \\
\hline
\end{tabular}
\end{table}

\begin{table}[p]
\caption{Superconducting state physical properties of samples of
\PrOsRuSb, determined from specific heat data. $T_c$ is the
superconducting transition temperature, $\Delta C$ is the jump in
$C(T)$ at $T_{c}$, $n$ is the exponent of the power-law fit below
$T_{c}$, $\gamma^{\rm e}_{\rm s}$ is the electronic specific heat
coefficient in the superconducting state ($\gamma^{\rm p}_{\rm s}
= 0$ for the power law fits), and $\Delta_{\rm e}$ is the
parameter in the exponential fit below $T_{c}$ that is
proportional to the energy gap.  The errors in the parameters were
determined by allowing $\Theta_{D}$ to vary by $\pm10$ K within
the fits (see text and Ref.~\cite{Frederick05} for details), with
the exception of $T_{c}$ and $\Delta C/T_{c}$ (the errors for
$\Delta C/T_{c}$ are represented graphically in
Fig.~\ref{DeltaCPhase}).}\label{Table2}
\bigskip
\scriptsize
\begin{tabular}{|c||cc|cc|cccccc|}
\hline
 & \multicolumn{2}{c|}{upper transition} & \multicolumn{2}{c|}{lower transition}
 & \multicolumn{6}{c|}{full transition} \\
$x$ & $T_{c1}$ & $\Delta C_{1}/T_{c1}$ & $T_{c2}$ & $\Delta
C_{2}/T_{c2}$ & $T_{c}$ & $\Delta C/T_{c}$ & $\Delta
C/\gamma{}T_{c}$ & $n$ & $\gamma_{\rm s}^{\rm e}$ & $\Delta_{\rm e}$ \\
 & (K) & (mJ/mol K$^2$) & (K) & (mJ/mol K$^2$) & (K) & (mJ/mol K$^2$) &
 & & (mJ/mol K$^2$) & (K) \\
\hline
0    & 1.84 & 505 & 1.73 & 661 & 1.77 & 1029 & 1.76 & 2.27$\pm$0.01 & 93.2$\pm$3.1 & 3.97$\pm$0.07 \\
0.01 & 1.79 & 393 & 1.65 & 258 & 1.73 & 608  & 0.74 & 2.56$\pm$0.01 & 65.1$\pm$0.6 & 3.73$\pm$0.02 \\
0.02 & 1.77 & 345 & 1.62 & 249 & 1.71 & 549  & 0.65 & 2.53$\pm$0.01 & 82.1$\pm$0.5 & 3.93$\pm$0.01 \\
0.04 & 1.72 & 337 & 1.58 & 145 & 1.68 & 481  & 0.66 & 2.56$\pm$0.02 & 63.6$\pm$1.3 & 3.77$\pm$0.03 \\
0.05 & ---  & --- & ---  & --- & 1.63 & 327  & 0.42 & 2.71$\pm$0.01 & 55.2$\pm$0.7 & 3.79$\pm$0.01 \\
\hline
\end{tabular}
\end{table}

\end{document}